\documentclass[review,number,sort&compress]{elsarticle}

\usepackage{lineno}

\usepackage{amssymb}

\journal{Nuclear Instruments and Methods in Physics Research Section A}

\begin{document}

\begin{frontmatter}



\title{A Functional Form for Liquid Scintillator Pulse Shapes}


\author{F.Q.L. Friesen\corref{cor1}}
\author{C.R. Howell}
\address{Department of Physics, Duke University and the Triangle Universities Nuclear Laboratory (TUNL).
Durham, North Carolina 27708, USA}

\cortext[cor1]{Corresponding author: fqf@tunl.duke.edu}

\begin{abstract}
Digitization of detector signals enables analysis of the original waveform to extract timing, particle identification, and energy deposition information. Here we present the use of analytical functions based on sigmoids to model and fit such pulse shapes from liquid organic scintillators, though the method should also be applicable to other detector systems. Neutron and gamma interactions in NE213 detectors were digitized from the phototube anode and fit using a sigmoid-based function. The acuity of the fit in extracting timing information and performing neutron-gamma pulse-shape discrimination are presented and discussed. 

\end{abstract}


\end{frontmatter}


\section{Introduction}
\label{Introduction}
Organic liquid scintillators are the dominant technology used in experiments requiring fast neutron detection. A wide variety of scintillators have been developed offering different light output characteristics. A common feature is a rapid rise in amplitude, followed by a slow exponential return to baseline. The decay tail may consist of multiple exponential components depending on the formulation of the liquid and particle being detected. For example, NE213 and the equivalent formulations of EJ301 and BC501A, are designed for fast neutron counting with good pulse-shape discrimination \cite{ADAMS1978459} (PSD) capability for distinguishing gamma from neutron interactions in the detector \cite{COMRIE201543}. The sensitivity of the decay time of the emitted light to the type of particle interaction in the liquid is due to the dependence on the mechanism of excitation. Gamma rays generally scatter from electrons, which then preferentially cause the liquid to fluoresce. Neutrons generally scatter from protons in the organic molecules, and the resulting excitations produce phosphorescence in addition to fluorescence \cite{birks1964}. The lifetime associated with phosphorescence is typically orders of magnitude larger, and pairs of the corresponding triplet states recombine with each other leading to a final delayed de-excitation by fluorescence \cite{Schuster2016}.

It is desirable to have a functional form to generically represent neutron and gamma pulse shapes. In addition to describing the decay, the function should accurately represent the rising edge of the pulse. It should have few free parameters, each with a clear physical interpretation. Uncertainties in the quantities of interest are then directly evaluated by the fit. With these features, it is possible to perform a fit on a given waveform, alter a parameter in a physically interpretable way, and regenerate a different but realistic signal. In this way, neutron and gamma waveforms with arbitrary timing can be quickly generated for simulating events in the same data format produced by the experimental apparatus. This ultimately enables processing of simulation output and measurements using identical code. 

A frequently used expression for the pulse shape has the form $L = A(e^{-\theta(t-t_0)} - e^{-\tau_f(t-t_0)}) + B(e^{-\theta(t-t_0)} - e^{-\tau_s(t-t_0)})$ \cite{MARRONE2002299}. While the number of free parameters is reasonable, obtaining reliable convergence during the fit process can be challenging. This difficulty increases if the location of the peak in time isn't well constrained. The initial (concave up) rise from baseline is also not modeled by this expression. In sufficiently fast scintillators, this initial rise can be non-negligible. In this work we examine the use of sigmoids for describing detector pulse shapes. They capture the exponential behavior with the benefit of remaining bounded. The proposed function is the product of several sigmoids representing the key features of the detector signals, e.g., the rise and decay times. 

The widespread use of digitizers, supported by cheap commodity storage, have made it common practice to keep detailed records of every critical signal in an experiment with nanosecond precision. With this capability, PSD analysis no longer requires dedicated hardware, e.g., using fixed width integration windows \cite{COMRIE201543}. More detailed signal analyses using fits becomes possible. 

\begin{figure}[btp]
\begin{center}
\includegraphics[width=5.0in]{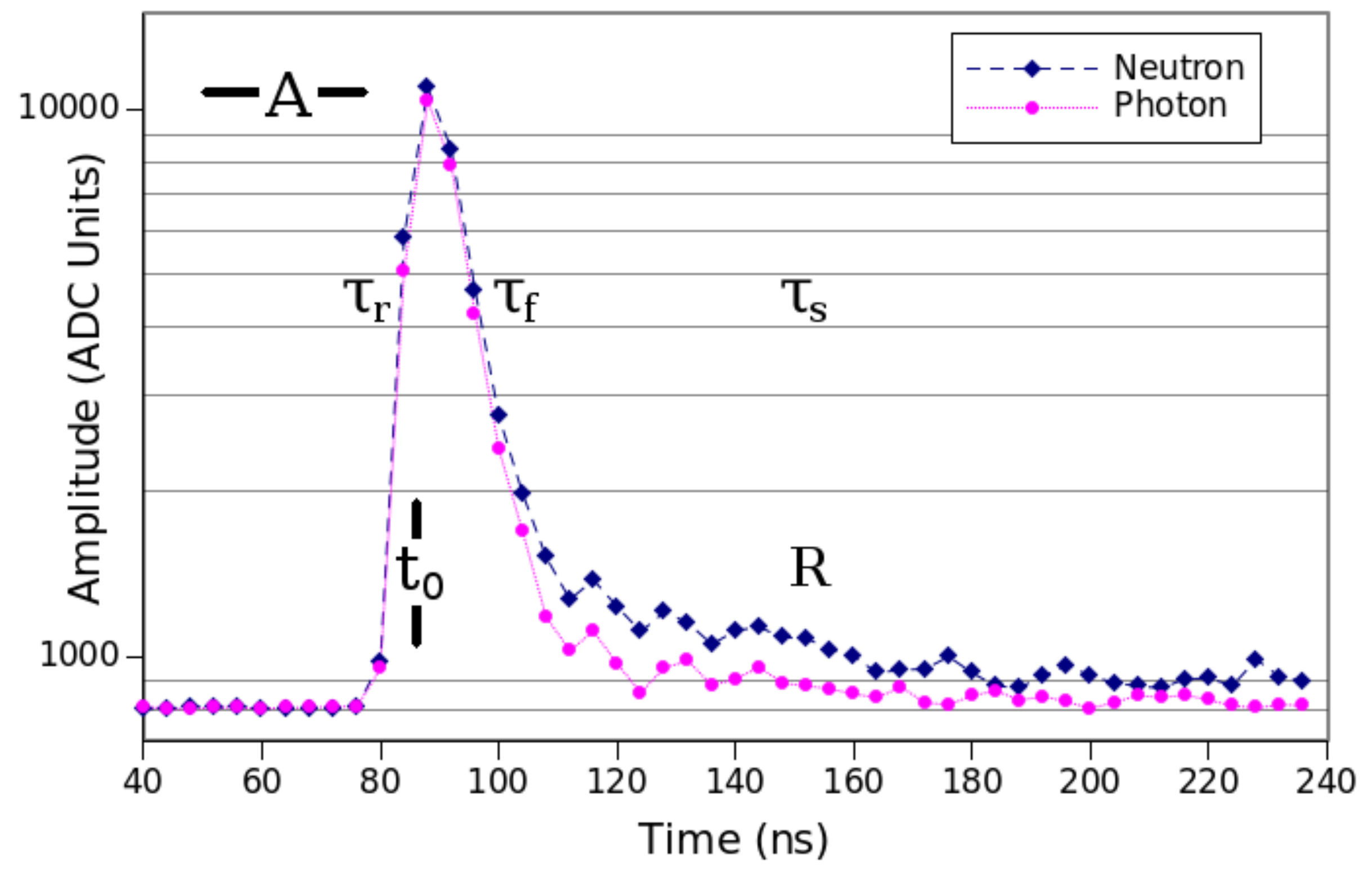}
\end{center}
\caption[Raw Waveforms]{Example raw waveforms from neutron and photon interactions with similar amplitudes. Labels for the time constants are positioned to indicate the regions dominated by each variable. The key difference between the interactions is the prevalence of the slow decay component $\tau_s$ as parameterized by $R$.}
\label{fig:neutron_gamma_raw}
\end{figure}

The traditional, time-tested PSD analysis method uses the charge comparison technique by computing the ratio of charge collected from the detector across two different fixed windows which correspond to the total charge in the pulse and the charge in the fast component (i.e., the total charge minus the charge in the tail)\cite{ADAMS1978459} \cite{HELTSLEY1988441}. A drawback of this method is that it can be difficult to find a pair of windows which work robustly over a wide range of pulse heights. Even though decay times are independent of amplitude, the pulse height of a signal determines how quickly the tail will become comparable to noise. This means that large signals are best characterized with longer windows; the opposite is true for small signals. This can lead to undesirable trade-offs in experiments when the detector threshold is low and neutrons and gammas need to be reliably distinguished across a large dynamic range. 

This work was performed in support of $^3He(\gamma,pn)$ measurements at TUNL. The functions are primarily used to generate digitizer-like data from the output of GEANT4 simulations. The functions can also be used in signal analysis by using them to fit the raw digitizer data. However, the use of these functions in data analysis would come at the expense of computation time relative to other methods.

\section{Experimental Setup}
\label{Experimental Setup}
Neutron detector efficiency calibration data were obtained using the $^3H(p,n)$ reaction at a polar angle of 136.4$^\circ$ relative to the incident proton beam axis. The digitized PMT signals were fitted using the proposed function. A proton beam was incident on a tritiated titanium foil target, producing neutrons at \mbox{0.8 MeV}, \mbox{1.0 MeV}, \mbox{1.2 MeV}, \mbox{1.4 MeV}, and \mbox{1.6 MeV}. The 1.6 MeV data are presented here, as they provide the largest range of pulse heights. The backward angle was chosen by requiring a higher proton beam energy than used for forward angle neutron production to reduce the energy spread of the neutrons. 

\begin{figure}[htbp]
\begin{center}
\includegraphics[width=5in]{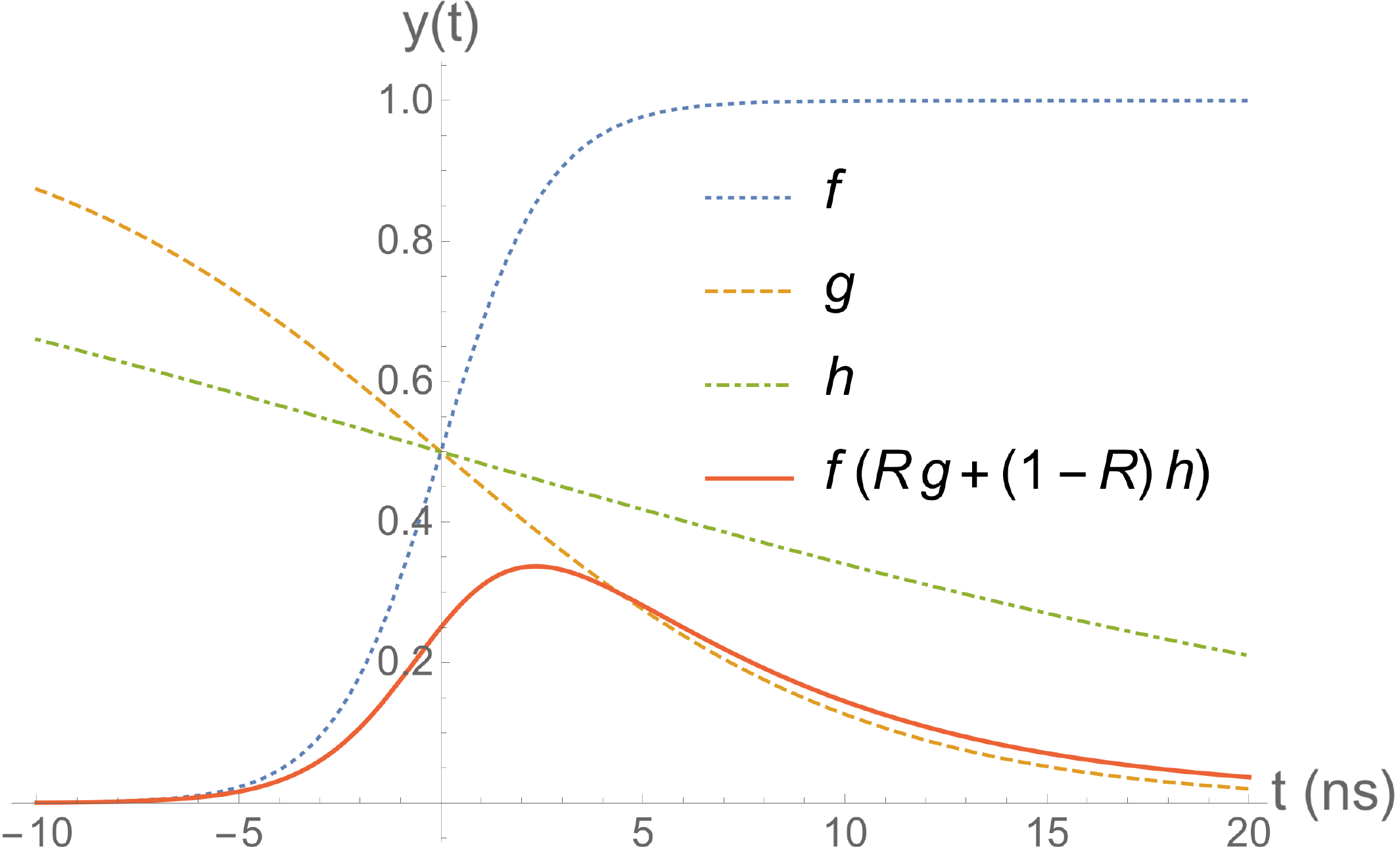}
\end{center}
\caption[Construction of neutron pulse shape]{The red solid line shows a representative neutron pulse shape as constructed from constituent sigmoids with $R=0.914$. The slow decay contribution is non-negligible.}
\label{fig:neutron_sigmoids}
\end{figure}

The detectors used in this study were filled with BC501A and coupled to Hamamatsu R1250 photomultiplier tubes. The interior of the liquid scintillator cells were cylindrical with diameter \mbox{12.7 cm} and depth of \mbox{5.1 cm}. The PMT anodes were connected directly to sis3316 digitizers. The digitizers have 12 bit resolution with a 4 ns sampling period. Waveforms were recorded for 240 ns such that the first 76 ns consisted of baseline as shown in Fig.\ref{fig:neutron_gamma_raw}. The full duration of sampled data was used for fitting.

\section{Pulse-Shape Analysis}
\label{Pulse-Shape Analysis}

For BC501A, the relative weighting of the importance of the two decay rates is a free parameter in the fit. This ratio $R$ is the fraction of decay light output produced by the fast mechanism. The expectation is that long decay times should be much less prominent in gamma events than neutron events. There are 6 independent parameters to fit for a PMT signal from a scintillation event in BC501A. The first 3 are fixed for a given PMT and scintillator combination, the remaining 3 are fit to each individual waveform (see Fig.\ref{fig:neutron_gamma_raw}):
\begin{itemize}
\item $\tau_{r}$ = Rise time ($\sim$1 ns)
\item $\tau_{f}$ = Fast decay time ($\sim$5 ns) representing fluorescence
\item $\tau_{s}$ = Slow decay time ($\sim$50 ns) representing return from phosphorescence
\item $R$ = Relative weighting of decay times $0 \leq R \leq 1$, e.g. $R$=1 means that the amplitude of the slow decay component is zero
\item $A$ = Pulse amplitude 
\item $t_0$ = Pulse arrival time 
\end{itemize}

\begin{equation}
\label{eq:time_resp}
y(t) = Af(t)[ R g(t) + (1-R) h(t) ] 
\end{equation}
 
where

\begin{equation}
\label{eq:time_resp_f}
f(t) = \frac{1}{e^{-\frac{t - t_0}{ \tau_{r} }}+1} 
\end{equation}

\begin{equation}
\label{eq:time_resp_g}
g(t) = 1 - \frac{1}{e^{-\frac{t - t_0}{ \tau_{f} }}+1} = \frac{1}{e^{\frac{t - t_0}{ \tau_{f} }}+1}
\end{equation}

\begin{equation}
\label{eq:time_resp_h}
h(t) = 1 - \frac{1}{e^{-\frac{t - t_0}{ \tau_{s} }}+1} = \frac{1}{e^{\frac{t - t_0}{ \tau_{s} }}+1}
\end{equation}

The constituent functions of Eq.\ref{eq:time_resp} are standard logistic functions \cite{seggern2006}. Those representing the decay of the pulse have been subtracted from unity. Eq.\ref{eq:time_resp_f} describes the early exponential rise from baseline at the start of the pulse and part of the round-off as the amplitude approaches its maximum. Note that there is no time offset between functions $f$, $g$, and $h$. This means that when the rise time is much smaller than the decay times, then function f has negligible effect on the tail.

Example waveforms are shown in Fig.\ref{fig:neutron_gamma_raw} with labels indicating the features determined by each parameter. Waveforms were prepared for the fit by first ensuring that the baseline level averages to 0. The subtraction was performed on an event-by-event basis. The curve fit function from the scipy.optimize \cite{jones2001} python library was used to perform the fit with bounded variables. Examples of the functions described by Eqs.\ref{eq:time_resp}-\ref{eq:time_resp_h} are shown in Fig.\ref{fig:neutron_sigmoids} for experimentally determined parameter values. The following bounds (in ns) were used to help the fit converge quickly:
\begin{equation}
\label{eq:bounds}
0 \le R \le 1 \quad
A \ge 0 \quad
0 \le \tau_r \le 6 \quad
0 \le \tau_f \le 24 \quad
40 \le \tau_s \le 100 \quad
\end{equation}

\begin{figure}[tbh]
\begin{center}
\includegraphics[width=5.5in]{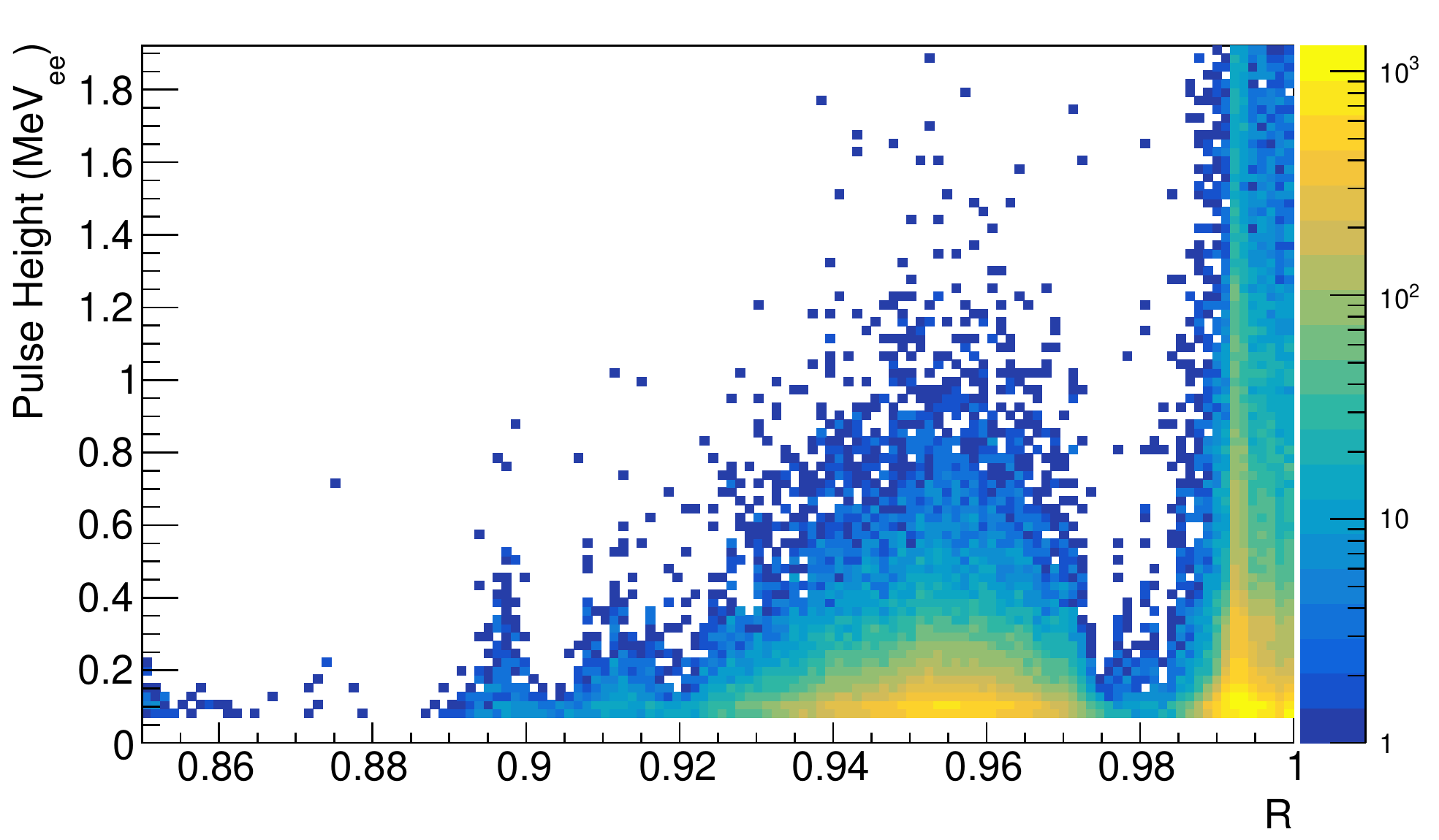}
\end{center}
\caption[R parameter vs Pulse Height]{The $R$ parameter plotted vs. waveform pulse height with $\tau_f$ and $\tau_s$ held constant during the fits to template generated samples. The lower R values are more neutron-like. The lower pulse height threshold corresponds to 60 keV$_{ee}$ ($1/8$ $^{137}$Cs edge). This histogram was accumulated with an incident neutron energy of 1.6 MeV.}
\label{fig:r_vs_ph}
\end{figure}

The waveforms taken from data were divided into three categories: templates, general samples, and template generated samples. The templates were those with a pulse height equivalent to the $^{137}$Cs edge or larger. Traditional PSD works very well in this region, so gamma pulses and neutron pulses can be unambiguously tagged by type before being fit. The general samples were drawn from the same pool of waveforms as the templates, but with a much lower pulse height cut ($1/8$ of the $^{137}$Cs edge). General sample waveforms were used to evaluate the timing capability of the fit. Template generated samples were constructed from templates by scaling down a template waveform amplitude to the obtain the desired pulse height, and then adding Gaussian noise to reproduce the original detector noise level (the pulse height spectrum was set to duplicate that of the general samples). The template generated samples were used to evaluate PSD capability.

\begin{figure}[tbp]
\begin{center}
\includegraphics[width=4in]{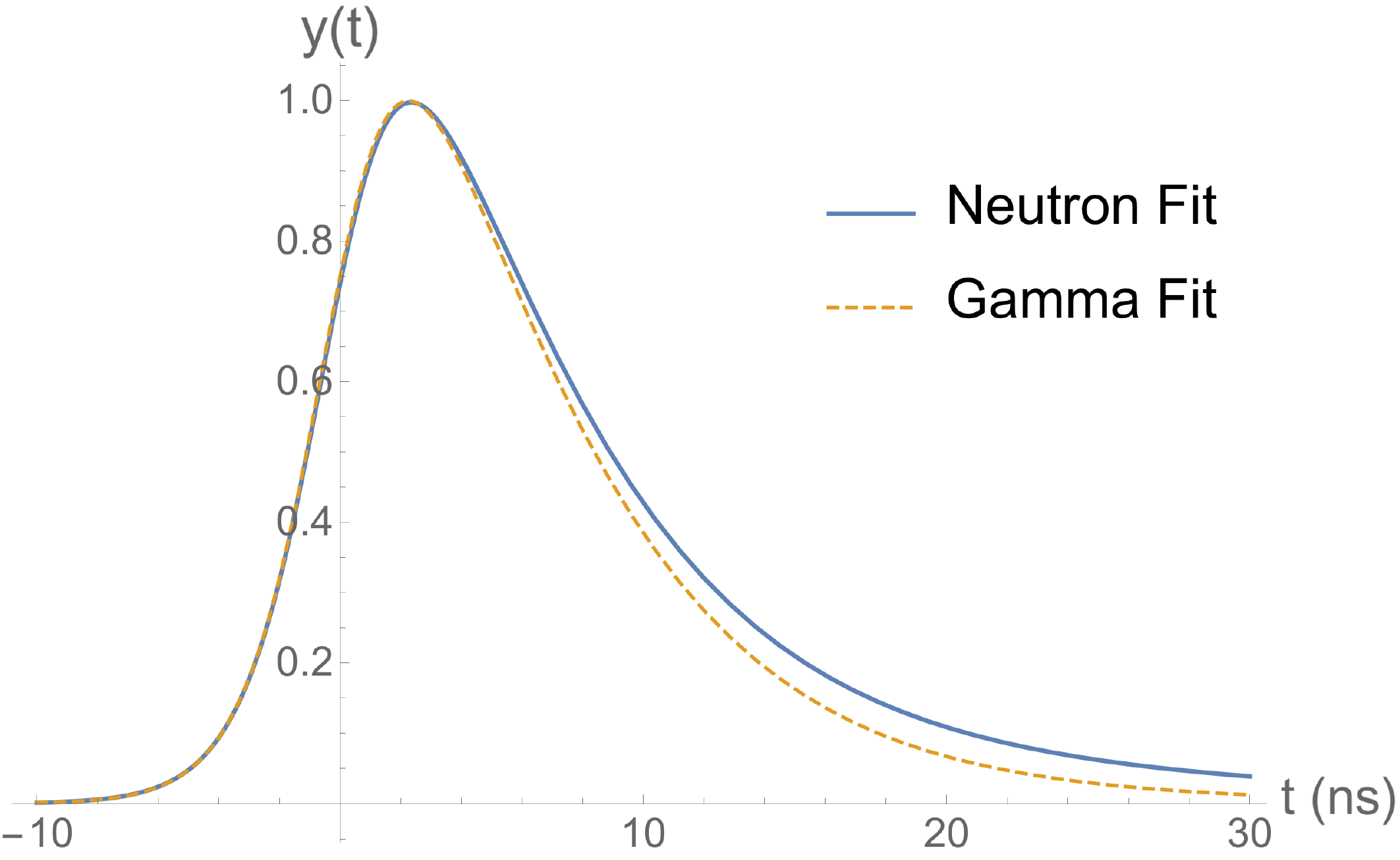}
\end{center}
\caption[Representative Pulse Shapes]{Example of normalized mean gamma and neutron pulse shapes obtained from fits to template waveforms.}
\label{fig:neutron_gamma_sigmoid_templates}
\end{figure}

The templates were fit individually using the bounds shown in Eq.\ref{eq:bounds}. The mean $\tau_f$ and $\tau_s$ values from neutron template fits were used to fix these parameters for all subsequent fits (all $\gamma$ waveforms, general samples, and template generated samples). The parameters $R$, $A$, $\tau_r$ and $t_0$ were left free. The $R$ can be interpreted as a PSD parameter (Fig.\ref{fig:r_vs_ph}). The PSD capability of $R$ is improved if $\tau_r$ is kept as a free parameter in the fit. Representative pulse shapes for neutron and photon events were generated as shown in Fig.\ref{fig:neutron_gamma_sigmoid_templates}.

\begin{figure}[tb]
\begin{center}
\includegraphics[width=5in]{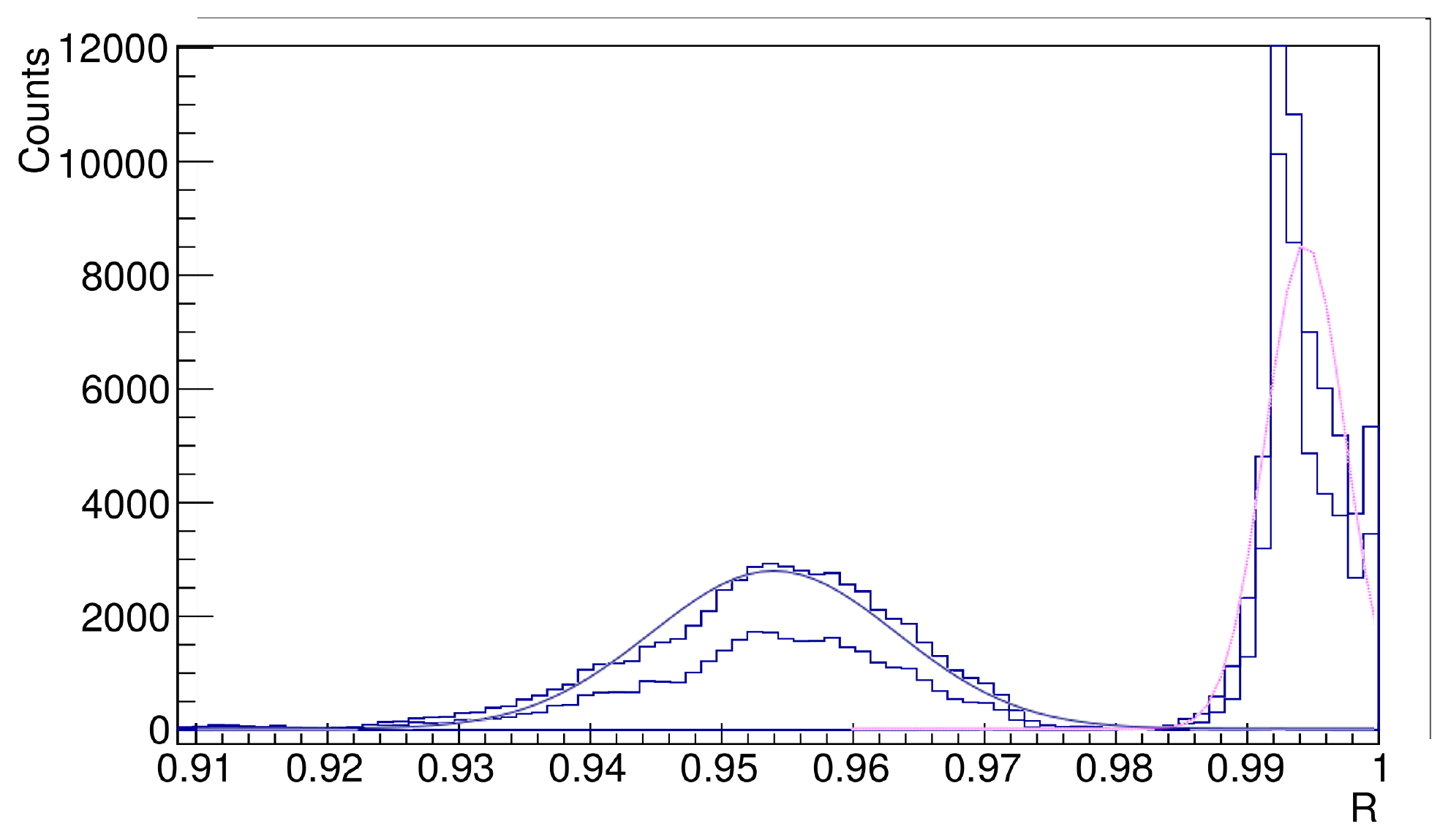}
\end{center}
\caption[PSD at Different Thresholds]{Histogram of fit (PSD) parameter R from a pool of template generated waveforms. The upper lines are for a neutron pulse height cut of 1/8 Cs, the lower lines are at 1/4 Cs. The left distributions show known neutron inputs. The right distributions show known gamma inputs. The gaussian fits used to calculate the FOM at $1/8$ $^{137}$Cs are also shown.}
\label{fig:run3590_R_psd_phcut_8th_4th}
\end{figure}

\begin{figure}[thb]
\begin{center}
\includegraphics[width=5in]{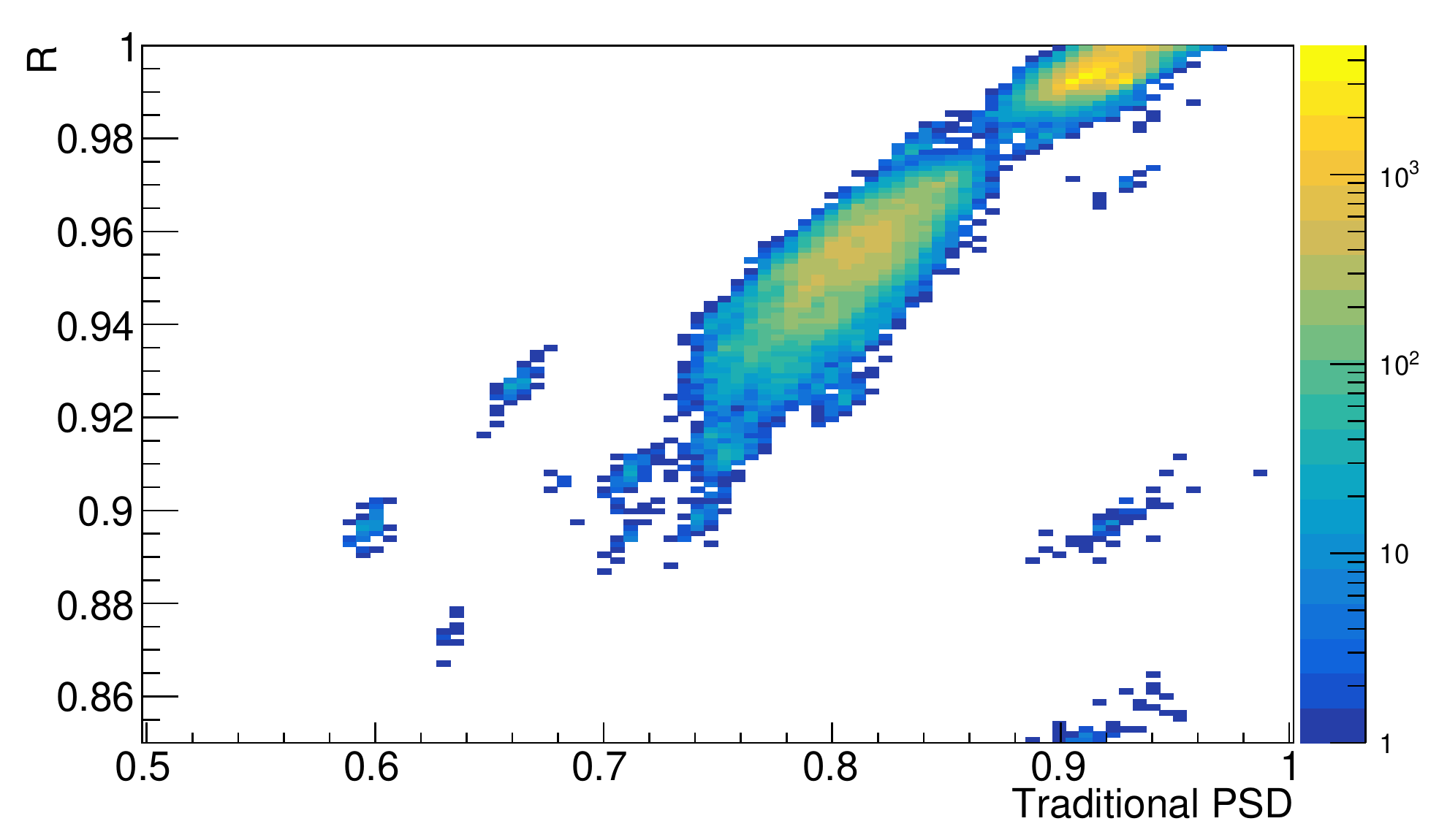}
\end{center}
\caption[PSD Uncertainty]{The PSD parameter calculated by the window integration method plotted against the $R$ fit parameter for the same data. The peak in the upper right hand corner corresponds to $\gamma$ events.}
\label{fig:psd_trad_vs_R_psd}
\end{figure}

Finally, the template scaling method provides a means to test PSD performance. Templates with known identities are scaled down and added to background noise to simulate lower pulse height events. Pools of realistic waveforms with low pulse-heights can be produced from templates with known identities. Using these generated events, the pulse identification capability can be quantitatively evaluated. PSD results for two sets of waveforms at different pulse height thresholds are shown in Fig.\ref{fig:run3590_R_psd_phcut_8th_4th}. The $R$ value from the fit is compared to a traditional PSD calculation in Fig.\ref{fig:psd_trad_vs_R_psd}. The traditional PSD parameter was calculated as the quotient of two baseline-subtracted waveform integrals. They extend to 20 and 4 samples past the leading edge, respectively (i.e., 172 ns and 108 ns in Fig.\ref{fig:neutron_gamma_raw}). There is good agreement over the classification of the vast majority of waveforms.  

\begin{table}[thb]
\caption{Neutron/$\gamma$ discrimination capability as a function of pulse height threshold. The discriminator threshold for $R$ was set to 0.98 for this analysis. Increasing it reduces rejection of true neutron events at the expense of mis-identifying more $\gamma$ events as neutrons. The uncertainties are statistical.}
\begin{tabular}[c]{lllll}
\hline
Threshold & Correctly     & Correctly           & $\#$ n above & $\#$ $\gamma$ above \\
(MeVee)   & Identified n  & Identified $\gamma$ & Threshold  & Threshold  \\
\hline
0.0598    & 0.997$\pm$0.006 & 0.998$\pm$0.006       & 6.0x10$^4$      & 6.0x10$^4$      \\
0.12      & 0.997$\pm$0.008 & 0.998$\pm$0.009       & 3.4x10$^4$      & 4.3x10$^4$      \\
0.239     & 1.00$\pm$0.01  & 1.00$\pm$0.03        & 1.1x10$^4$      & 2.5x10$^4$      \\
0.478     & 1.00$\pm$0.03  & 1.0$\pm$0.1          & 2.1x10$^3$      & 1.2x10$^4$     
\end{tabular}
\label{tab:psd_quality_threshold}
\end{table}

The result of fixing the discrimination threshold on $R$ at 0.98 is shown in table \ref{tab:psd_quality_threshold}. The table was generated from a pool of 1.2 x10$^5$ tagged waveforms (generated by template scaling) with a 1/8 Cs (60 keV$_{ee}$) pulse height threshold limit, where half are neutrons and the other half are photons. At this threshold, 2 $\gamma$-ray signals leak through the PSD cut for every 1000 photons detected. For every 1000 incident neutrons, 3 are misidentified as photons. The reduction in total number of neutron and gamma events as the threshold is increased corresponds to the reduced detection efficiency. The values and uncertainties of parameters obtained from fitting the template signals are given in Table \ref{tab:param_vals}. A figure of merit (FOM) value was also calculated for the $R$ discrimination parameter by fitting the neutron and gamma peaks with gaussian functions and using the definition FOM$=(\mu_\gamma-\mu_n)/(\sigma_\gamma+\sigma_n)$ where $\sigma$ refers to the FWHM of each gaussian \cite{BOSE1988487}. The traditional charge comparison PSD method had FOM values of 1.34 and 1.43 for pulse height thesholds of 1/8 and 1/4 Cs, respectively. The FOM values for the $R$ fit method were 1.39 and 1.52 calculated on the same data.

\begin{table}[thb]
\caption{Mean values and standard deviation of the parameters obtained by fitting templates. The pool consisted of 12,000 waveforms with an equal number of neutron and photon events. The uncertainty in $t_0$ is given as an estimate of the time resolution of this method because the absolute value of $t_0$ depends on the source of the particle. The $\tau_f$ and $\tau_s$ values for photons were constrained to be the same as the mean values obtained from neutron template fits. Therefore, these fixed parameters do not have uncertainties for photons.}
\begin{tabular}{lll}
\hline
Parameter     & Neutrons          & $\gamma$-rays     \\
\hline
$\tau_r$ (ns) & 1.46 $\pm$ 0.29   & 1.37 $\pm$ 0.26   \\
$\tau_f$ (ns) & 5.45 $\pm$ 0.63   & 5.45    \\
$\tau_s$ (ns) & 52 $\pm$ 11   & 52   \\
$R$           & 0.949 $\pm$ 0.035 & 0.992 $\pm$ 0.050 \\
$t_0$ (ns)    & $\pm$ 0.17        & $\pm$ 0.11       
\end{tabular}
\label{tab:param_vals}
\end{table}

\section{Timing Analysis} 
\label{Timing Analysis}

\begin{figure}[thb]
\begin{center}
\includegraphics[width=5in]{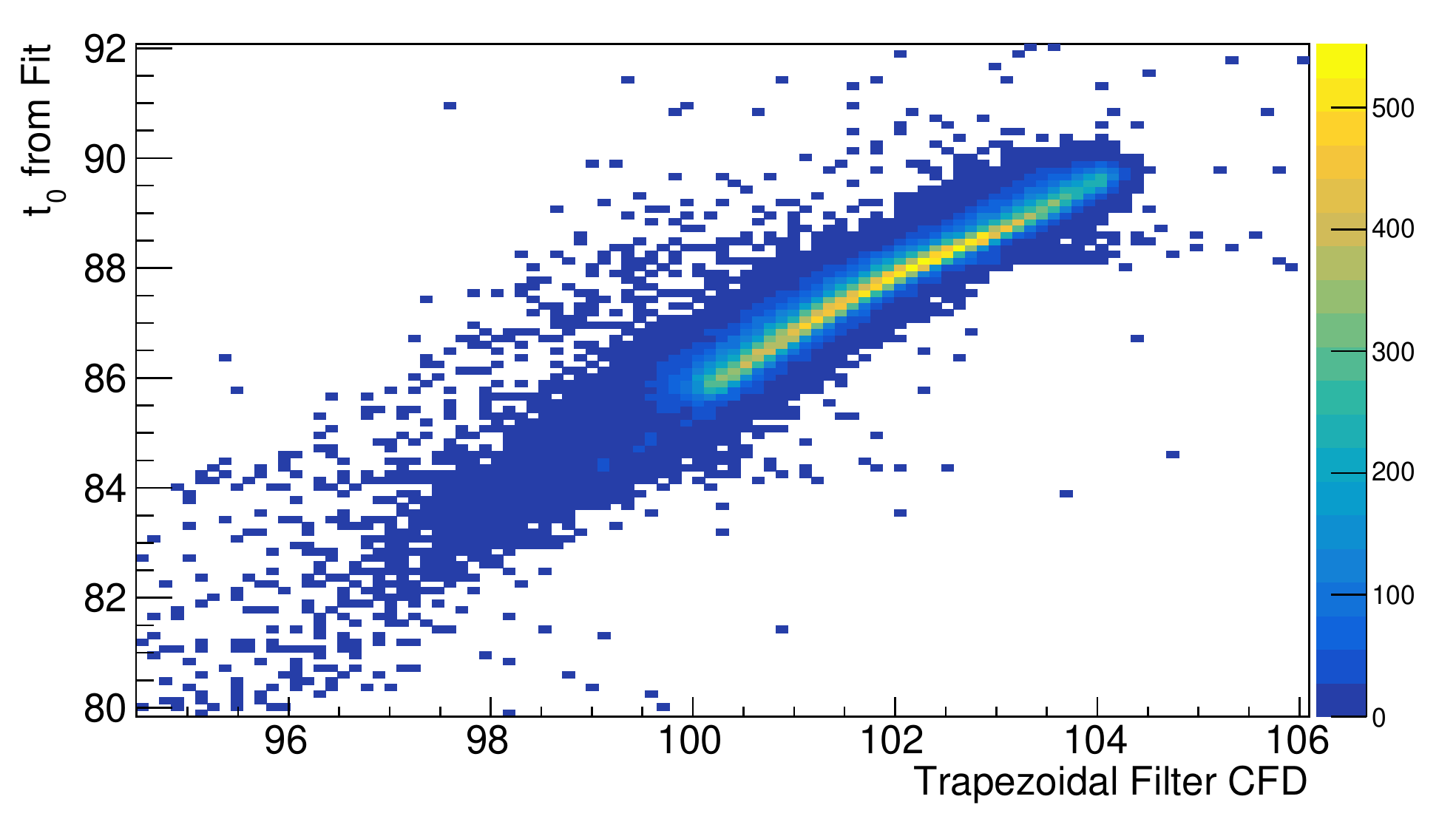}
\end{center}
\caption[Trapezoidal filter CFD vs t_0 from fit]{Trapezoidal filter CFD vs $t_0$ from the sigmoid-based fit applied to general samples. The pulse height threshold corresponds to 60 keV$_{ee}$ ($1/8$ $^{137}$Cs edge). Slices passing through the bin with the most counts (102.6 ns, 88.1 ns) were used to compare timing resolution.}
\label{fig:scfd_vs_t0}
\end{figure}

The $t_0$ fit parameter was compared with a software constant fraction discriminator \cite{cfdref} (CFD) value computed with a trapezoidal filter. The data sampling period was 4 ns. The peak and gap times were both 2 samples long (\mbox{8 ns}), and the time was interpolated on the falling edge of the filter at $1/2$ of the maximum output. The relationship between the two, shown in Fig.\ref{fig:scfd_vs_t0}, indicates that the raw $t_0$ value from the fit closely follows the CFD value independent of event type or amplitude. Slices taken either perpendicular or parallel to the CFD axis in Fig.\ref{fig:scfd_vs_t0} corresponding to $85.6 \le t_0 \le 89.6$ ns (100 to 104 ns on the CFD axis) were used to compare timing resolution. These are experimental data, so different arrival times have different physical interpretations, and different slices should not be directly combined. The standard deviation of the points in the vertical slices from the fit is 0.33 ns within the range, compared to a standard deviation of 0.35 ns as determined with the CFD method (horizontal slices passing through the same points).

\section{Conclusions} 
A sigmoid-based functional form was developed for representing pulse shapes from liquid scintillation detectors. The rise time, amplitude, time offset, and decay components are represented in a physically interpretable way. Contributions from different decay times can be reliably identified by the fit, thereby enabling PSD analysis of each waveform. The bounded nature of sigmoids simplifies construction of plausible functions, and the function presented here can be further generalized to allow for more decay components as needed. The function parametrized in Eqs.\ref{eq:time_resp}-\ref{eq:time_resp_h} can be used to generate signals to realistically represent pulse shapes from liquid scintillator detectors. Waveforms associated with BC501A can be generated using values given in table \ref{tab:param_vals}. Other sigmoids could also be explored for constructing the fit function.



\bibliographystyle{elsarticle-num} 
\bibliography{scint_pulse_shapes_bibtex}

\begin{thebibliography}{10}
\expandafter\ifx\csname url\endcsname\relax
  \def\url#1{\texttt{#1}}\fi
\expandafter\ifx\csname urlprefix\endcsname\relax\def\urlprefix{URL }\fi
\expandafter\ifx\csname href\endcsname\relax
  \def\href#1#2{#2} \def\path#1{#1}\fi

\bibitem{ADAMS1978459}
J.~Adams, G.~White, A versatile pulse shape discriminator for charged particle
  separation and its application to fast neutron time-of-flight spectroscopy,
  Nuclear Instruments and Methods 156~(3) (1978) 459 -- 476.

\bibitem{COMRIE201543}
A.~Comrie, A.~Buffler, F.~Smit, H.~Wörtche, Digital neutron/gamma
  discrimination with an organic scintillator at energies between {1MeV and
  100MeV}, Nuclear Instruments and Methods in Physics Research Section A:
  Accelerators, Spectrometers, Detectors and Associated Equipment 772 (2015) 43
  -- 49.

\bibitem{birks1964}
J.~Birks, The theory and practice of scintillation counting, International
  series of monographs on electronics and instrumentation 27.

\bibitem{Schuster2016}
P.~Schuster, E.~Brubaker, Investigating the anisotropic scintillation response
  in anthracene through neutron, gamma-ray, and muon measurements, IEEE
  Transactions on Nuclear Science 63~(3) (2016) 1942--1954.

\bibitem{MARRONE2002299}
S.~Marrone, D.~Cano-Ott, N.~Colonna, C.~Domingo, F.~Gramegna, E.~Gonzalez,
  F.~Gunsing, M.~Heil, F.~Käppeler, P.~Mastinu, P.~Milazzo, T.~Papaevangelou,
  P.~Pavlopoulos, R.~Plag, R.~Reifarth, G.~Tagliente, J.~Tain, K.~Wisshak,
  Pulse shape analysis of liquid scintillators for neutron studies, Nuclear
  Instruments and Methods in Physics Research Section A: Accelerators,
  Spectrometers, Detectors and Associated Equipment 490~(1) (2002) 299 -- 307.

\bibitem{HELTSLEY1988441}
J.~Heltsley, L.~Brandon, A.~Galonsky, L.~Heilbronn, B.~Remington, S.~Langer,
  A.~V. Molen, J.~Yurkon, J.~Kasagi, Particle identification via pulse-shape
  discrimination with a charge-integrating adc, Nuclear Instruments and Methods
  in Physics Research Section A: Accelerators, Spectrometers, Detectors and
  Associated Equipment 263~(2) (1988) 441 -- 445.

\bibitem{seggern2006}
D.~H. von Seggern, CRC Standard Curves and Surfaces with Mathematica, Chapman
  and Hall/CRC, 2016.

\bibitem{jones2001}
E.~Jones, et~al., \href{http://www.scipy.org/}{{SciPy}: Open source scientific
  tools for {Python}} (2001).
\newline\urlprefix\url{http://www.scipy.org/}

\bibitem{BOSE1988487}
S.~Bose, M.~Chatterjee, B.~Sinha, R.~Bhattacharyya, Neutron-gamma
  discrimination based on leading edge shape measurement, Nuclear Instruments
  and Methods in Physics Research Section A: Accelerators, Spectrometers,
  Detectors and Associated Equipment 270~(2) (1988) 487 -- 491.

\bibitem{cfdref}
M.~Nakhostin, Signal Processing for Radiation Detectors, John Wiley and Sons,
  2017.

\end{thebibliography}


%
%
%

\end{document}